# Methodology For the Evaluation of Critical Components of the Scatterable Radiation Monitor Under Radiation Fields

Matthew Niichel*, Stylianos Chatzidakis[†]

*2dLt, United States Space Force, Purdue University, 500 Central Dr, West Lafayette, IN 47907, mniichel@purdue.edu.
[†]PhD, Purdue University, 500 Central Dr, West Lafayette, IN 47907, schatzid@purdue.edu

## INTRODUCTION

One of the persisting challenges associated with the peaceful use of the atom is monitoring the subsequent radiation. For example, nuclear power plants need environmental monitoring to ensure public trust. While this may be conducted by the frequent monitoring of low-cost thermo-luminescent dosimeters (TLD) placed in concentric rings around the plant, there have been several solutions for a modernized real-time digital alternative to the TLD. The Scatterable Radiation Monitoring (SCRAM) device [1] is a STM32 microprocessor-based design aiming at the wireless transmission of environmental radiation data. At this point, the performance of the critical features on SCRAM while exposed to gamma and neutron fields is unknown. This paper highlights the methods used to evaluate the SCRAM device while placed in Purdue University's sub-critical pile.

## BACKGROUND

The STM 32 microcontroller is an "off-the-shelf" hardware, the processor of choice for hobbyists and industry alike. Although radiation damage to electronics is a well-studied area, irradiation of the STM 32 chip lacks publicly available data when compared to other electronics. Of such, two studies exist focusing solely on the irradiation of the STM 32.

The first considered a Sr-90 source and a STM32 chip irradiated at 2.5 GBq. The purpose was to simulate a low Earth orbit space mission at approximately 10 years, the team concluded the dose to be 3 krad. The study primarily focused on the universal synchronous/asynchronous serial receiver/transmitter where they have support for STM32 failure near 47 krad of total ionizing dose. [2] However, it should be mentioned that the damage inflicted by the beta decay from Sr-90 is significantly different from gamma and neutron radiation.

The second study considered a method for testing the bipolar transistors under fast neutrons which is more of interest to the SCRAM project due to the larger damaging effect. However, the report did not present any significant findings and only provided the methodology for measuring the voltage gain as a means of determining internal damage to the STM32. [3]

Unrelated to the STM32, another study considers a generic computer chip under irradiation. The primary focus was to classify the extent of damage caused to electronics by thermal neutrons. The team makes an analytical comparison of neutron damage in electronics based on damage in biological tissues. The results of their experiments and calculations state that a neutron flux of $1\times10^5$ (n/cm$^2$s) and a dwell time of 15 hours provide sufficient damage to become noticeable in the performance of a generic computer chip. [4]

## CHARACTERIZING THE SUB-CRITICAL PILE

In December of 2023, the Purdue sub-critical pile was rearranged to accommodate an aluminum test bed for the use of testing electronics under gamma and neutron fields. The prior arrangement of the pile was well characterized by students, professors, and researchers alike. However, the placement of the test bed required another classification of the pile. Figure 1 displays the current layout of the pile, where green represents natural U-metal slugs, yellow is a PuBe source, red is an AmBe source, black is graphite, and white is a void or the test bed.

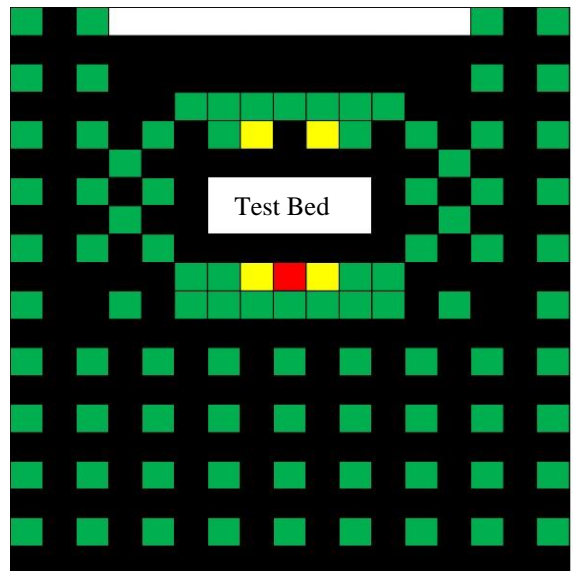

Fig. 1 Layout of the sub-critical pile with test bed.

The process for determining the neutron flux is defined in ASTM E262 *via* the irradiation of metal foils. While there are several options, gold is the most available. By placing the foils on the test bed in the pile, the neutron flux can be determined through the gamma decay of Au-198 created by the neutron conversion of Au-197. Equation (1) shows how to derive the neutron flux through the gold foils. Figure 2 displays the calculated neutron flux from the gold foil experiment.

$$\Phi = \frac{A}{N\sigma_a(1-e^{-\lambda t_a})e^{-\lambda t_w}} \quad (1)$$

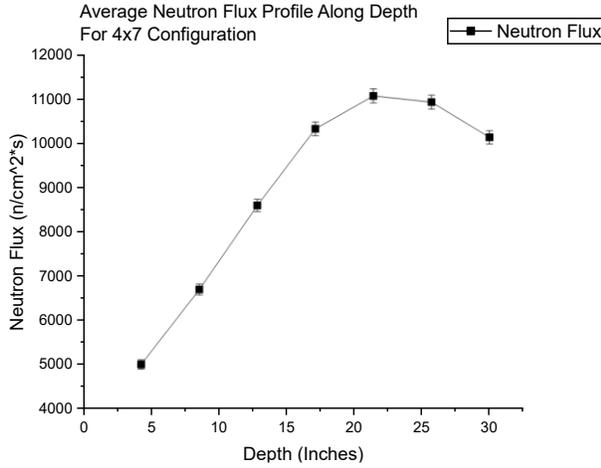

Fig. 2 Average neutron flux along the profile of the test bed using a 4x7 array of gold foils.

In addition to the total neutron flux, it is important to determine the ratio of fast flux to thermal flux within the pile as the two groups cause damage to electronics asymmetrically. A similar metal foil experiment is designed to calculate the fast neutron flux with the use of gold and cadmium where the flux of a cadmium-covered foil is compared to that of a bare gold foil. The ratio between the two identifies the epithermal and thermal fluxes. Equations (2)-(4) describe these relations. [5]

$$R = \frac{A_b}{A_c} \quad (2)$$

$$\Phi_{th} = \frac{A_b - A_c}{N\sigma_a(1-e^{-\lambda t_a})e^{-\lambda t_w}} \quad (3)$$

$$\Phi_f = \frac{\Phi_{th}\sigma_a}{(R-1)*1526b} * log\left(\frac{2.5\ MeV}{0.025\ eV}\right) \quad (4)$$

The average of the cadmium ratio R was found to be about 3 throughout the pile. This was conducted by taking the integral of both the bare activity and cadmium activities and then dividing the two. Then making use of equations (3) and (4) the flux profile can be derived for both energy groups as a function of distance within the test bed. The resulting thermal and fast flux profiles are shown in Figure 3.

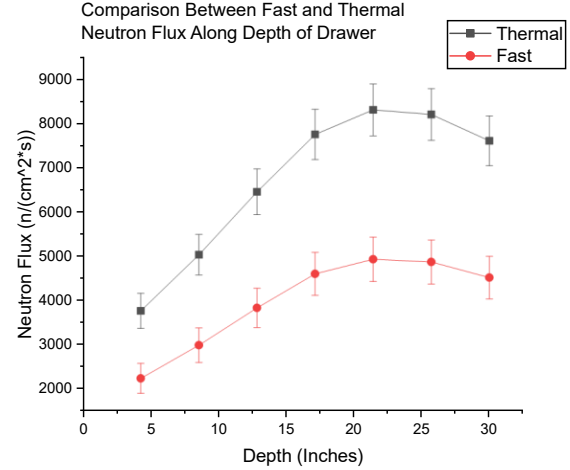

Fig. 3 Resulting fast and thermal flux profiles along the depth of the testbed drawer in the sub-critical pile.

**REMOTE MONITORING OF SCRAM**

An inherent issue with measuring the signals of the SCRAM with other electronics is the potential for damage to occur to the monitoring equipment. To prevent a malfunction of the non-target electronics, a reasonable distance must be placed between the monitoring setup and the pile.

One solution is making use of an Arduino Mega for collecting the signals of SCRAM. The device, then can be powered and monitored by ethernet following IEEE protocol 802.3af. This document ensures that voltages are within the standard of the Cat5e cable construction. Furthermore, in making use of Arduino, there needs to be certainty that the reported values are accurate. By using ethernet, there is the introduction of additional wire resistance and there is uncertainty that setting up the Arduino using the Analog inputs can measure voltages within an acceptable tolerance. [6]

Two experiments are designed to determine if the setup of the Arduino is within reporting tolerance. The SCRAM internal voltage is 3.3 Volts and the SiPM output voltage is 35 Volts. Given this data, a voltage divider is crafted to measure the two voltages, with each voltage having 3 channels. The additional channels, allow for coincidence measurements, minimizing any erratic behavior or noise in any one channel.

Both experiments made use of a high-voltage power supply, the Arduino was used to measure the output and then was compared to a Fluke 177. The Fluke in this experiment was considered to be accurate and the comparison standard. The first experiment considered voltage in 1Volt increments from 1 to 40. This would allow for curve fitting and the generation of a correction function to allow the Arduino to report the same as the Fluke. The second experiment focused on finding the voltage limit of the voltage divider by applying 10 Volt increments up to 100 volts. The results of the two experiments are displayed in Figures 4 and 5 respectively. The latter provides evidence that the limit for a divider made of 1 MΩ and 100 kΩ is 50 Volts.

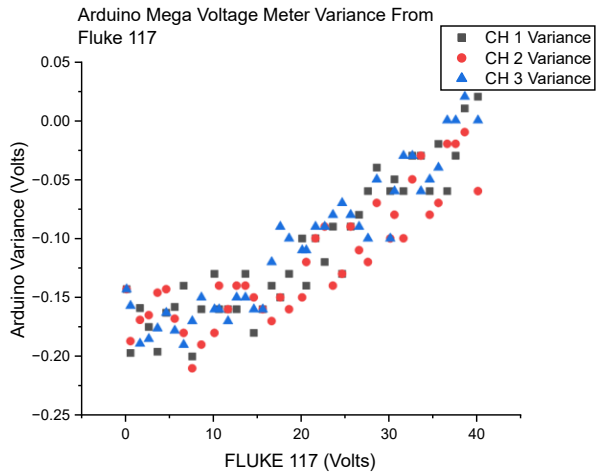

Fig 4. Scatter plot of three channels measured from high voltage power supply relative to the Fluke 117 measurement.

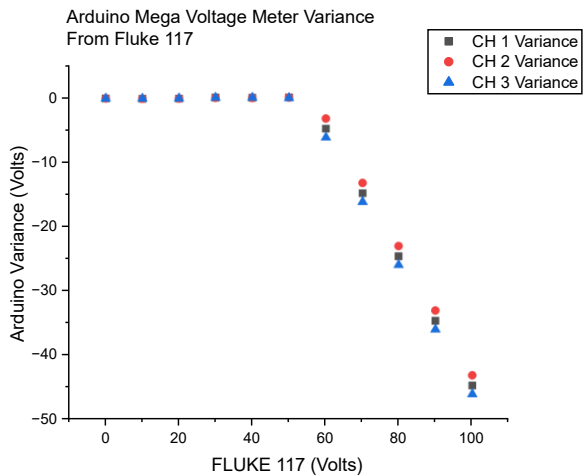

Fig. 5. Scatter plot of three channels highlighting the limits of the voltage divider.

**EVALUATING THE EFFECTS OF RADIATION**

The chips placed in the pile for testing are not a complete version of the SCRAM. The components selected for performance evaluation were done so by subjective means on the importance of overall SCRAM functionality. The 3.3 Volt signal was selected because it evaluates the low voltage regulator. The 35 Volt signal was selected as it evaluates the output of the high voltage output MOSFET supplying bias to the SiPM. Finally, the analog-digital converter (ADC) was selected because this is used to send and receive serial commands. If any of these signals were to vary significantly compared to an established baseline, then it becomes indicative of a component failure. More broadly, the importance of these components on the scale of the SCRAM device would be indicative of an overall system failure. As a result of the selection of the low voltage regulator, high voltage power supply, and ADC combined with the need for coincidence measurements, a total of 7 signals are collected. The 3.3 Volt and the 35 Volt receive 3 channels each and the ADC receives 1 channel. The ADC receives a single channel because the data transmitted is serial. Any significant variation in the serial data will result in an incorrect value. Thereby indicating an error.

A baseline of the signals produced by the SCRAM is collected by applying power to the device over ethernet. Then the Arduino monitors the 7 signals and reports them using serial communication. A figure of the three channels for both voltages is displayed with the respective average. An example of these figures produced of the 3.3 V, 34 V, and ADC outputs are displayed in figures 6-8 respectively.

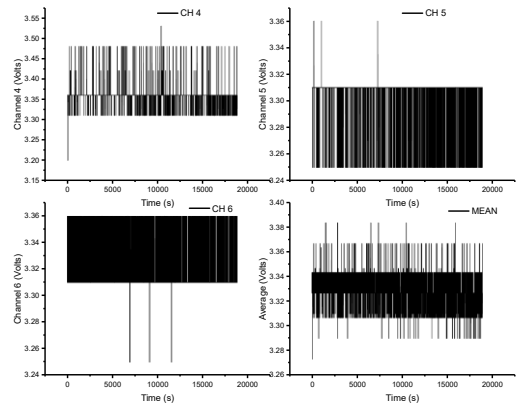

Fig. 6. Example output of the baseline 3.3 V signal with average.

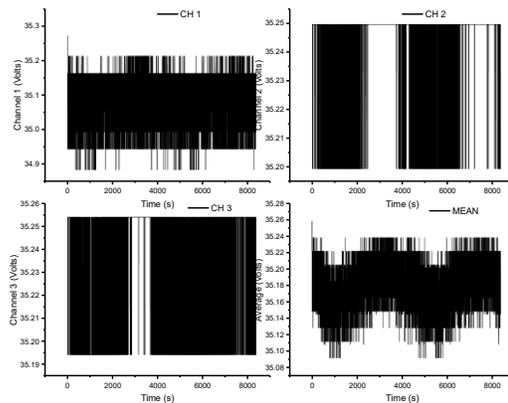

Fig. 7. Example output of the baseline 34 V signal with average.

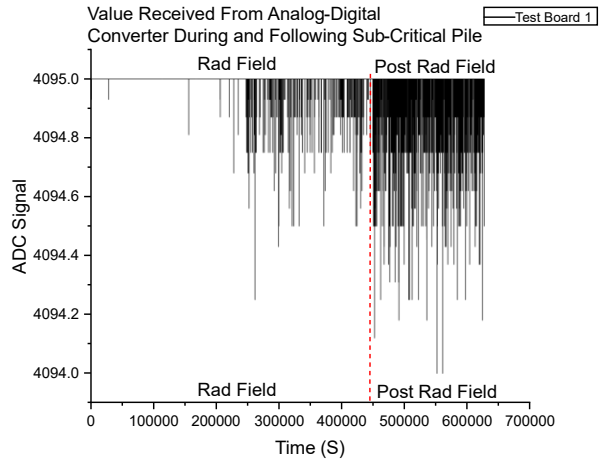

Fig. 8. Example output of the baseline ADC value.

Following the successful collection of baseline signals for several hours/days the SCRAM chip can be placed within the pile for 7-10 days to achieve approximately the same neutron dose as the general electronics discussed in the earlier section. Measurements will be taken concurrently with time spent in the pile and following the test chip's removal. A comparison between the signals will be performed making use of equations (5)-(7).

$$\sigma = \frac{N_e}{\phi} \quad (5)$$

$$MWTF = (\sigma - \phi - execution\ time)^{-1} \quad (6)$$

$$MTBF = \frac{total\ working\ time - total\ breakdown\ time}{number\ of\ breakdowns} \quad (7)$$

## CONCLUSION

While the results of testing SCRAM in the sub-critical pile have yet to be collected, characterizing the pile and conducting experiments on the Arduino provides support for the methodology by which the components will be tested and evaluated when subjected to radiation.

A firm understanding of the radiation field in which the test subject is placed will allow for proper classification of damage to the electronics if any occurs at all. By examining the Arduino voltage divider, support is provided that the maximum working voltage for the divider is 50 Volts and that the variation from the Fluke 117 standard can be identified for a correction of the recorded signals. Finally, the collection of multiple signals to include three channels for each voltage reading allows for an understanding of the SCRAM device's proper operation.

There is much work to be conducted on the actual evaluation of the electronic components. However, there is evidence to support the methodology for successfully testing and evaluating the critical components of the SCRAM device.

## NOTATION

$A_b$ = Bare gold foil activity (Bq).
$A_c$ = Cadmium-covered gold foil activity (Bq).
$\lambda$ = Decay constant for Au-198 (s).
N = Number of particles.
MWTF = Mean work to failure (s$^{-1}$)
MTBF = Mean time between failures (s).
$N_e$ = Number of errors.
$t_a$ = Activation time (min).
$t_w$ = Cooling time (min).
R = Cadmium ratio.
$\sigma$ = bit per cross-section.
$\sigma_a$ = Microscopic cross-section for absorption for Au-197 (cm$^{-2}$).
$\phi$ = Total neutron flux (n/cm$^2$s).
$\Phi_{th}$ = Thermal neutron flux (n/cm$^2$s).
$\Phi_f$ = Epithermal neutron flux (n/cm$^2$s).

## ACKNOWLEDGMENTS
This research was performed using the Purdue College of Engineering and Purdue Military Research Institute funding.
## DISCLAIMER
*The views expressed in this article are those of the author and do not reflect the official policy or position of the United States Air Force, Department of Defense, or the U.S. Government.*

## REFERENCES

1. D. FOBAR, "Scatterable radiation monitoring," *Defense Threat Reduction Agency West Point, NY.* (2023).
1. P. MADLE, "STM32H7 Radiation Test Report," *DOC-01251-01: Open Source Satellite,* (2021)
2. V. I. BUTIN, A. V. BUTINA, "Bipolar transitor application ffor on-line neutron fluence" *Russia: 2017 International Multi-Conference on Engineering, Computer and Information Sciences (SIBIRCON).* (2017).
3. T. F. KRAEMER SARZI SARTORI, " Assessment of Radiation Effects on Attitude Estimation Processing for Autonomous Things," *IEEE Transactions on Nuclear Science, 69,* 1610-1617. (2022).
4. S. SHALBI, W. SALLEH, F. IDRIS, " Neutron measurement at the thermal column of the Malaysian Triga Reactor Mark II," *Materials Science and Engineering, 298, (2017).*
5. IEEE. "IEEEE Standard for Information Technology - Telecommunications and Information Exchange Between Systems - Local and Metropolitan Area Networks - Specific Requirements - Part 3: Carrier Sense Multiple Access with Collision Detection (CSMA/CD) Access Method and Physical Layer Specifications - Data Terminal Equipment (DTE) Power Via Media Dependent Interface (MDI) " *IEEE 802.3af,* (2003).